\documentclass[%
twocolumn,
superscriptaddress,
nofootinbib,
 amsmath,amssymb,
 prc
 aps,
]{revtex4-1}

\usepackage{romanbar}
\usepackage{xcolor}
\usepackage{graphicx}
\usepackage{dcolumn}
\usepackage{bm}
\usepackage{SIunits}
\usepackage{array}

\begin{document}

\title{Search for Boosted Dark Matter in COSINE-100}
\author{G.~Adhikari}
\affiliation{Department of Physics and Wright Laboratory, Yale University, New Haven, CT 06520, USA}
\author{N.~Carlin}
\affiliation{Physics Institute, University of S\~{a}o Paulo, 05508-090, S\~{a}o Paulo, Brazil}
\author{J.~J.~Choi}
\affiliation{Department of Physics and Astronomy, Seoul National University, Seoul 08826, Republic of Korea} 
\affiliation{Center for Underground Physics, Institute for Basic Science (IBS), Daejeon 34126, Republic of Korea}
\author{S.~Choi}
\affiliation{Department of Physics and Astronomy, Seoul National University, Seoul 08826, Republic of Korea} 
\author{A.~C.~Ezeribe}
\affiliation{Department of Physics and Astronomy, University of Sheffield, Sheffield S3 7RH, United Kingdom}
\author{L.~E.~Fran{\c c}a}
\affiliation{Physics Institute, University of S\~{a}o Paulo, 05508-090, S\~{a}o Paulo, Brazil}
\author{C.~Ha}
\affiliation{Department of Physics, Chung-Ang University, Seoul 06973, Republic of Korea}
\author{I.~S.~Hahn}
\affiliation{Department of Science Education, Ewha Womans University, Seoul 03760, Republic of Korea} 
\affiliation{Center for Exotic Nuclear Studies, Institute for Basic Science (IBS), Daejeon 34126, Republic of Korea}
\affiliation{IBS School, University of Science and Technology (UST), Daejeon 34113, Republic of Korea}
\author{S.~J.~Hollick}
\affiliation{Department of Physics and Wright Laboratory, Yale University, New Haven, CT 06520, USA}
\author{E.~J.~Jeon}
\affiliation{Center for Underground Physics, Institute for Basic Science (IBS), Daejeon 34126, Republic of Korea}
\author{J.~H.~Jo}
\affiliation{Department of Physics and Wright Laboratory, Yale University, New Haven, CT 06520, USA}
\author{H.~W.~Joo}
\affiliation{Department of Physics and Astronomy, Seoul National University, Seoul 08826, Republic of Korea} 
\author{W.~G.~Kang}
\affiliation{Center for Underground Physics, Institute for Basic Science (IBS), Daejeon 34126, Republic of Korea}
\author{M.~Kauer}
\affiliation{Department of Physics and Wisconsin IceCube Particle Astrophysics Center, University of Wisconsin-Madison, Madison, WI 53706, USA}
\author{B.~H.~Kim}
\affiliation{Center for Underground Physics, Institute for Basic Science (IBS), Daejeon 34126, Republic of Korea}
\author{H.~J.~Kim}
\affiliation{Department of Physics, Kyungpook National University, Daegu 41566, Republic of Korea}
\author{J.~Kim}
\affiliation{Department of Physics, Chung-Ang University, Seoul 06973, Republic of Korea}
\author{K.~W.~Kim}
\email{kwkim@ibs.re.kr}
\affiliation{Center for Underground Physics, Institute for Basic Science (IBS), Daejeon 34126, Republic of Korea}
\author{S.~H.~Kim}
\affiliation{Center for Underground Physics, Institute for Basic Science (IBS), Daejeon 34126, Republic of Korea}
\author{S.~K.~Kim}
\affiliation{Department of Physics and Astronomy, Seoul National University, Seoul 08826, Republic of Korea}
\author{W.~K.~Kim}
\affiliation{IBS School, University of Science and Technology (UST), Daejeon 34113, Republic of Korea}
\affiliation{Center for Underground Physics, Institute for Basic Science (IBS), Daejeon 34126, Republic of Korea}
\author{Y.~D.~Kim}
\affiliation{Center for Underground Physics, Institute for Basic Science (IBS), Daejeon 34126, Republic of Korea}
\affiliation{Department of Physics, Sejong University, Seoul 05006, Republic of Korea}
\affiliation{IBS School, University of Science and Technology (UST), Daejeon 34113, Republic of Korea}
\author{Y.~H.~Kim}
\affiliation{Center for Underground Physics, Institute for Basic Science (IBS), Daejeon 34126, Republic of Korea}
\affiliation{Korea Research Institute of Standards and Science, Daejeon 34113, Republic of Korea}
\affiliation{IBS School, University of Science and Technology (UST), Daejeon 34113, Republic of Korea}
\author{Y.~J.~Ko}
\affiliation{Center for Underground Physics, Institute for Basic Science (IBS), Daejeon 34126, Republic of Korea}
\author{D.~H.~Lee}
\affiliation{Department of Physics, Kyungpook National University, Daegu 41566, Republic of Korea}
\author{E.~K.~Lee}
\affiliation{Center for Underground Physics, Institute for Basic Science (IBS), Daejeon 34126, Republic of Korea}
\author{H.~Lee}
\affiliation{IBS School, University of Science and Technology (UST), Daejeon 34113, Republic of Korea}
\affiliation{Center for Underground Physics, Institute for Basic Science (IBS), Daejeon 34126, Republic of Korea}
\author{H.~S.~Lee}
\email{hyunsulee@ibs.re.kr}
\affiliation{Center for Underground Physics, Institute for Basic Science (IBS), Daejeon 34126, Republic of Korea}
\affiliation{IBS School, University of Science and Technology (UST), Daejeon 34113, Republic of Korea}
\author{H.~Y.~Lee}
\affiliation{Center for Underground Physics, Institute for Basic Science (IBS), Daejeon 34126, Republic of Korea}
\author{I.~S.~Lee}
\affiliation{Center for Underground Physics, Institute for Basic Science (IBS), Daejeon 34126, Republic of Korea}
\author{J.~Lee}
\affiliation{Center for Underground Physics, Institute for Basic Science (IBS), Daejeon 34126, Republic of Korea}
\author{J.~Y.~Lee}
\affiliation{Department of Physics, Kyungpook National University, Daegu 41566, Republic of Korea}
\author{M.~H.~Lee}
\affiliation{Center for Underground Physics, Institute for Basic Science (IBS), Daejeon 34126, Republic of Korea}
\affiliation{IBS School, University of Science and Technology (UST), Daejeon 34113, Republic of Korea}
\author{S.~H.~Lee}
\affiliation{IBS School, University of Science and Technology (UST), Daejeon 34113, Republic of Korea}
\affiliation{Center for Underground Physics, Institute for Basic Science (IBS), Daejeon 34126, Republic of Korea}
\author{S.~M.~Lee}
\affiliation{Department of Physics and Astronomy, Seoul National University, Seoul 08826, Republic of Korea} 
\author{Y.~J.~Lee}
\affiliation{Department of Physics, Chung-Ang University, Seoul 06973, Republic of Korea}
\author{D.~S.~Leonard}
\affiliation{Center for Underground Physics, Institute for Basic Science (IBS), Daejeon 34126, Republic of Korea}
\author{N.~T.~Luan}
\affiliation{Department of Physics, Kyungpook National University, Daegu 41566, Republic of Korea}
\author{B.~B.~Manzato}
\affiliation{Physics Institute, University of S\~{a}o Paulo, 05508-090, S\~{a}o Paulo, Brazil}
\author{R.~H.~Maruyama}
\affiliation{Department of Physics and Wright Laboratory, Yale University, New Haven, CT 06520, USA}
\author{R.~J.~Neal}
\affiliation{Department of Physics and Astronomy, University of Sheffield, Sheffield S3 7RH, United Kingdom}
\author{J.~A.~Nikkel}
\affiliation{Department of Physics and Wright Laboratory, Yale University, New Haven, CT 06520, USA}
\author{S.~L.~Olsen}
\affiliation{Center for Underground Physics, Institute for Basic Science (IBS), Daejeon 34126, Republic of Korea}
\author{B.~J.~Park}
\affiliation{IBS School, University of Science and Technology (UST), Daejeon 34113, Republic of Korea}
\affiliation{Center for Underground Physics, Institute for Basic Science (IBS), Daejeon 34126, Republic of Korea}
\author{H.~K.~Park}
\affiliation{Department of Accelerator Science, Korea University, Sejong 30019, Republic of Korea}
\author{H.~S.~Park}
\affiliation{Korea Research Institute of Standards and Science, Daejeon 34113, Republic of Korea}
\author{K.~S.~Park}
\affiliation{Center for Underground Physics, Institute for Basic Science (IBS), Daejeon 34126, Republic of Korea}
\author{S.~D.~Park}
\affiliation{Department of Physics, Kyungpook National University, Daegu 41566, Republic of Korea}
\author{R.~L.~C.~Pitta}
\affiliation{Physics Institute, University of S\~{a}o Paulo, 05508-090, S\~{a}o Paulo, Brazil}
\author{H.~Prihtiadi}
\affiliation{Center for Underground Physics, Institute for Basic Science (IBS), Daejeon 34126, Republic of Korea}
\author{S.~J.~Ra}
\affiliation{Center for Underground Physics, Institute for Basic Science (IBS), Daejeon 34126, Republic of Korea}
\author{C.~Rott}
\affiliation{Department of Physics, Sungkyunkwan University, Suwon 16419, Republic of Korea}
\affiliation{Department of Physics and Astronomy, University of Utah, Salt Lake City, UT 84112, USA}
\author{K.~A.~Shin}
\affiliation{Center for Underground Physics, Institute for Basic Science (IBS), Daejeon 34126, Republic of Korea}
\author{D.~F.~F.~S. Cavalcante}
\affiliation{Physics Institute, University of S\~{a}o Paulo, 05508-090, S\~{a}o Paulo, Brazil}
\author{A.~Scarff}
\affiliation{Department of Physics and Astronomy, University of Sheffield, Sheffield S3 7RH, United Kingdom}
\author{N.~J.~C.~Spooner}
\affiliation{Department of Physics and Astronomy, University of Sheffield, Sheffield S3 7RH, United Kingdom}
\author{W.~G.~Thompson}
\affiliation{Department of Physics and Wright Laboratory, Yale University, New Haven, CT 06520, USA}
\author{L.~Yang}
\affiliation{Department of Physics, University of California San Diego, La Jolla, CA 92093, USA}
\author{G.~H.~Yu}
\affiliation{Department of Physics, Sungkyunkwan University, Suwon 16419, Republic of Korea}
\affiliation{Center for Underground Physics, Institute for Basic Science (IBS), Daejeon 34126, Republic of Korea}
\collaboration{COSINE-100 Collaboration}
\date{\today}

\begin{abstract}
		We search for energetic electron recoil signals induced by boosted dark matter (BDM) from the galactic center using the COSINE-100 array of NaI(Tl) crystal detectors at the Yangyang Underground Laboratory. 
		The signal would be an excess of events with energies above 4\,MeV over the well-understood background. Because no excess of events are observed in a 97.7\,kg$\cdot$years exposure, we set
		limits on BDM interactions under a variety of hypotheses. 
		Notably, we explored the dark photon parameter space, leading to  competitive limits compared to direct dark photon search experiments, particularly for dark photon masses below 4\,MeV and considering the invisible decay mode. Furthermore, by comparing our results with a previous BDM search conducted by the Super-Kamionkande experiment, we found that the COSINE-100 detector has advantages in searching for low-mass dark matter. 
		This analysis demonstrates the potential of the COSINE-100 detector to search for MeV electron recoil signals produced by the dark sector particle interactions. 
\end{abstract}
\maketitle

A number of astrophysical observations provide evidence that the dominant matter component of the Universe is not ordinary matter but rather non-baryonic dark matter~\cite{Clowe:2006eq,Planck:2018vyg}. 
Many searches for signs of dark matter have been pursued by direct detection experiments~\cite{Aprile:2018dbl,PandaX-4T:2021bab,LZ:2022ufs,XENONCollaboration:2023orw,Billard:2021uyg,Akerib:2022ort}, indirect detection experiments~\cite{Super-Kamiokande:2020sgt,ANTARES:2020leh,IceCube:2021xzo,HESS:2022ygk,Conrad:2017pms,Hooper:2018kfv}, and collider experiments~\cite{CMS:2018yiq,ATLAS:2021jbf,dmcol} without success~\cite{ParticleDataGroup:2022pth}.  
It motivates searches for alternative types of dark matter that produce substantially different signatures in detectors, such as light (mass) dark matter models that predict extremely low-energy signals~\cite{COSINE-100:2021poy,SENSEI:2021hcn,DarkSide:2022dhx,DAMIC-M:2023gxo} or relativistically boosted dark matter~(BDM) that produce more energetic signals~\cite{Super-Kamiokande:2017dch,COSINE-100:2018ged,Kim:2020ipj}.

A relativistic dark matter particle, $i.e.$ one that is boosted by interactions with cosmic-rays in the galaxy~\cite{Bringmann:2018cvk,Ema:2018bih,Cappiello:2019qsw,Xia:2021vbz,Jho:2021rmn,Das:2021lcr,Ghosh:2021vkt,Bardhan:2022bdg,Granelli:2022ysi,Alvey:2022pad} or produced by the decay~\cite{Bhattacharya:2014yha,Kopp:2015bfa,Heurtier:2019rkz} or annihilation~\cite{Alhazmi:2016qcs,Kim:2016zjx,Giudice:2017zke} of heavier dark sector particles, can deposit signals with energies that are above MeV in detectors. Since typical direct detection experiments search for low-energy nuclear recoil signals, scenarios for such energetic events have not been very well studied. 
Here we consider a model in which a BDM is produced by heavier dark matter particles~\cite{Agashe:2014yua,Kong:2014mia,Berger:2014sqa}.
It would require at least two species of dark matter particles, denoted by $\chi_0$ and $\chi_1$ for the heavier and lighter dark matter particles, respectively~\cite{Agashe:2014yua,Belanger:2011ww}. 
The first direct search for BDM from annihilations of heavy dark matter particles in the galactic center was performed with the Super-Kamiokande detector that searched for energetic electron recoil signals above 100\,MeV induced by BDM elastic scattering~\cite{Kachulis:2017nci}. With COSINE-100 data, we searched for the inelastic scattering of BDM (IBDM)~\cite{COSINE-100:2018ged} induced by the existence of another dark sector particle~\cite{Kim:2016zjx,Giudice:2017zke}. 
Recently, searches for cosmic-ray BDM interacting with protons in dark matter detectors with energies of a few keV~\cite{PandaX-II:2021kai,CDEX:2022fig}, as well as in neutrino detectors with energies between a few MeV~\cite{PROSPECT:2021awi} and a few GeV~\cite{Super-Kamiokande:2022ncz},  were performed.

In this Letter, we report on a search for BDM that elastically interacts with electrons in the NaI(Tl) crystals of the COSINE-100 detector. Such interactions would produce energetic electrons in the NaI(Tl) crystals. 
Our region of interest for the BDM interaction consists of an energy deposition above 4\,MeV, since radioactive $\gamma$ or $\beta$ particles primarily have energy less than 4\,MeV.

COSINE-100~\cite{Adhikari:2017esn} is composed of an array of eight ultra-pure NaI(Tl) crystals, each coupled to two photomultiplier tubes (PMTs). Due to high background levels and low light yields of three crystals, this analysis only uses data from five crystals, corresponding to an effective mass of 61.3\,kg~\cite{COSINE-100:2021xqn,COSINE-100:2021zqh}. The crystals are immersed in an active veto detector that is composed of 2,200\,L of linear alkylbenzene~(LAB)-based liquid scintillator~(LS)~\cite{Adhikari:2020asl}. The LS is contained within a shield comprising a 3\,cm thick layer of oxygen-free copper, a 20\,cm thick layer of lead, and an array of plastic scintillation counters for cosmic-ray muon tagging~\cite{Prihtiadi:2017inr,COSINE-100:2020jml}. 

We used data obtained between 21 October 2016 and 18 July 2018, corresponding to 1.7 years of effective live time, and a total exposure of 97.7\,kg$\cdot$years for this search.
The same dataset was already adopted for a precise understanding and modeling of the observed background between 1\,keV and 3\,MeV~\cite{cosinebg2}, as well as for a dark matter search that concentrated on the low-energy nuclear recoil spectrum~\cite{COSINE-100:2021xqn}. 

The COSINE-100 data acquisition system recorded two different signals from the crystal PMTs, covering a wide dynamic range from single-photoelectron to 4\,MeV high energy~\cite{COSINE-100:2018rxe}.  In addition to the low-energy (0--100\,keV) anode readout, the 5$^{\mathrm{th}}$ stage dynode readout  was recorded by 500\,MHz flash analog-to-digital converters (FADCs) for 8\,$\mu$s long waveforms. 
It provided sufficient energy resolution for events with energies between  50\,keV and 3\,MeV. 

\begin{figure*}[!htb] 
\begin{center}
		\begin{tabular}{cc}
\includegraphics[width=0.48\textwidth]{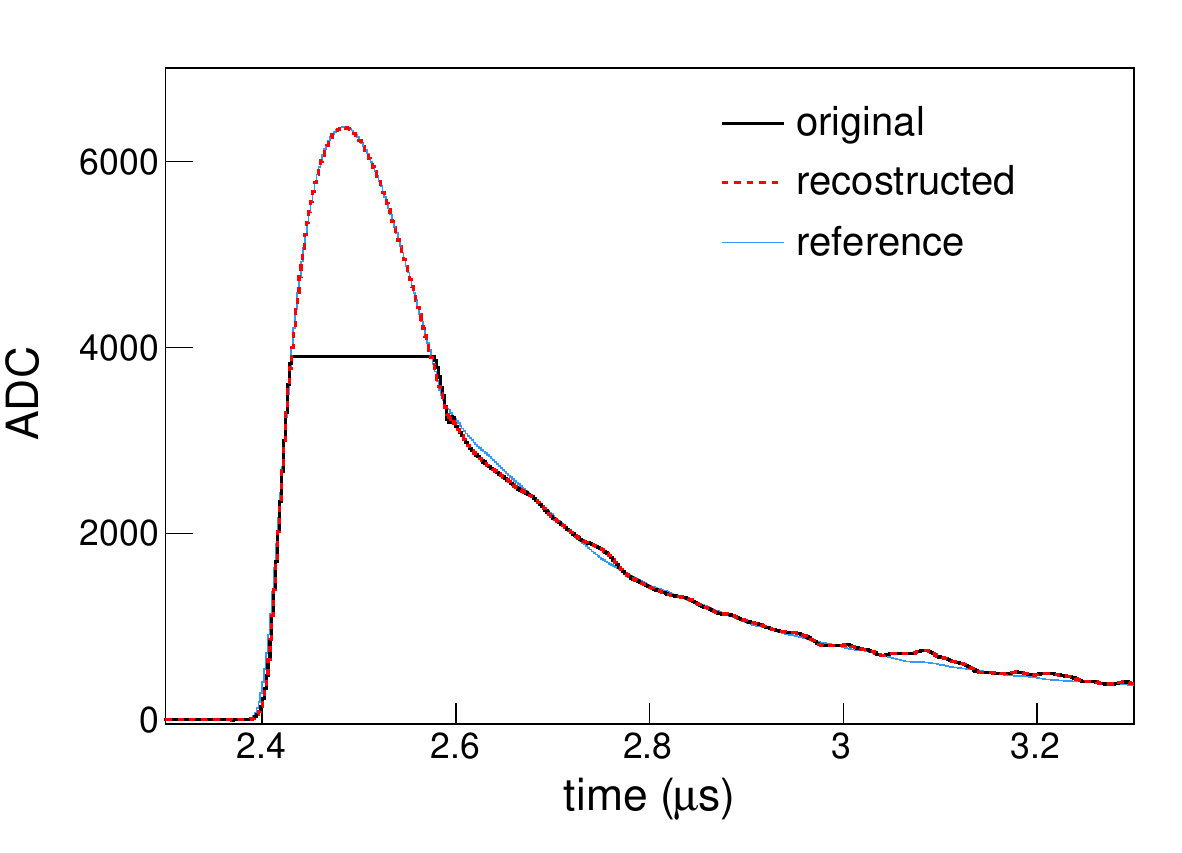} &
\includegraphics[width=0.48\textwidth]{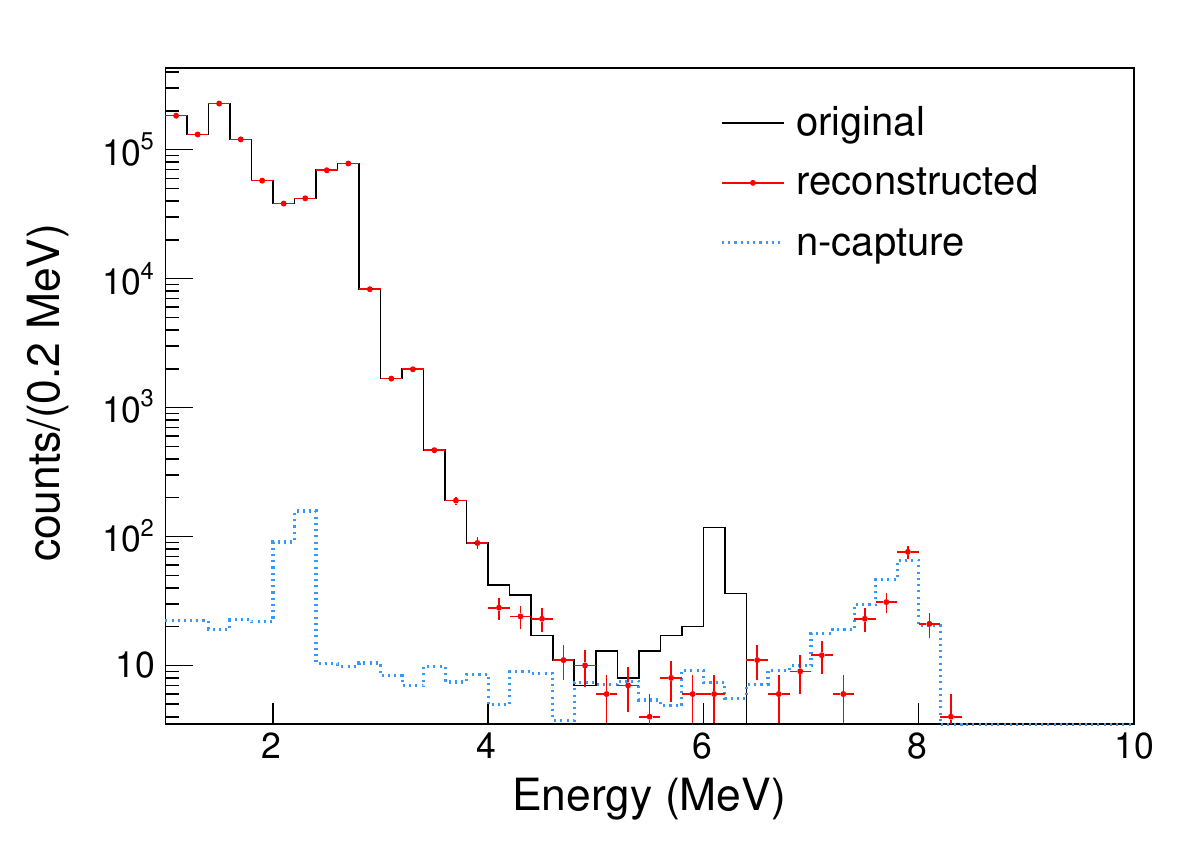}  \\
(a) Pulse shape & (b) Single-hit energy spectra \\
		\end{tabular}
\caption{
  (a) An example of a saturated event due to the limited dynamic range (12\,bit, 4096 for 2.5\,V) is presented as a black solid line. Reconstruction of the saturated event (red dashed line) is achieved by comparison with the template from unsaturated events (thin blue solid line). In this example, a 6.02\,MeV energy is reconstructed. (b) The measured energy spectra before (black-solid line) and after (red dots) reconstruction of the saturated events  are presented. The reconstructed energy spectrum is calibrated with 7.6\,MeV and 7.9\,MeV $\gamma$-rays from the neutron capture of iron and copper, respectively, as shown in the blue dotted line. 
}
\label{fig_event}
\end{center}
\end{figure*}

We have previously presented a background model for the COSINE-100 detectors that covered energies below 3\,MeV~\cite{cosinebg,cosinebg2}. 
However, events with energies greater than 4\,MeV were above the limit of the FADC dynamic range and suffered from a saturated, non-linear response. 
To address this issue, we developed an algorithm to detect the saturation of the recorded pulse and reconstruct the saturated event. 
A template from the unsaturated events at the 2--3\,MeV energy was compared to the saturated pulse, and the reconstruction at the saturated region was performed, as shown in Fig.~\ref{fig_event} (a). The original energy spectrum as well as the recovered energy spectrum are shown in Fig.~\ref{fig_event} (b). 

The energy scale above 4\,MeV is calibrated with  7.6\,MeV and 7.9\,MeV $\gamma$-rays from $^{56}$Fe and $^{63}$Cu, respectively, that are produced by thermal neutron capture in the steel supporter of the lead shield and the copper encapsulation of the NaI(Tl) crystals~\cite{Adhikari:2017esn}. Figure~\ref{fig_event} (b) shows the reconstructed energy spectrum of the single hit events, in which the spectrum above 6\,MeV is well described by the neutron capture events from Geant4~\cite{GEANT4:2002zbu}-based simulation.

Candidate events are selected if the reconstructed energy is greater than 4\,MeV with no coincident muon candidate tracks in the muon detector~\cite{Prihtiadi:2017inr}. We reject $\alpha$-induced events in the crystals using a pulse shape discrimination method~\cite{COSINE:2020egt}. 
Selected candidate events are sorted into two different categories: single-hit and multiple-hit events. 
A multiple-hit event has accompanying crystal signals with more than four photoelectrons or has a liquid scintillator signal above 80\,keV~\cite{Adhikari:2020asl}. 
A single-hit event is classified as one where the other detectors do not meet these criteria.
Although a BDM interaction with the NaI(Tl) crystal would generate an energetic single electron with energy between a few MeV and a few 100\,MeV~\cite{Kim:2016zjx,Giudice:2017zke}, this energetic electron could generate a number of Bremsstrahlung radiation-induced $\gamma$s that could convert and deposit energy in the other crystals or the LS. Therefore, we use both single and multiple-hit channels in this analysis.
Although the single-hit channel exhibits dominant sensitivity, the multiple-hit channel yields a slight enhancement in final sensitivity.

Four different categories of events contribute to the background above 4\,MeV. Internal or external $\beta/\gamma$ radiation induced by environmental radioactivities were well understood by the background modeling of the COSINE-100 detector for energy below 3\,MeV~\cite{cosinebg,cosinebg2}. We extended this model to energies above 4\,MeV. Here the main contribution is caused by internal $^{228}$Th decay, especially sequenced $^{212}$Bi ($\beta$-decay with a Q-value of 2.25\,MeV) and $^{212}$Po ($\alpha$-decay with a Q-value of 8.95\,MeV) decays with a 300\,ns half-life of $^{212}$Po. Because of the short half-life, $^{212}$Bi and $^{212}$Po events pile up in the 8\,$\mu$s event window. Based on their distinct pulse shapes, we can partially reject them, but their residual is the main contribution above 4\,MeV from environmental $\beta/\gamma$ radiation. 

Although the muon veto detector tags a coincident event with muons~\cite{COSINE-100:2020jml}, 2.14$\pm$0.21\% of the muons that transit the detector are mis-tagged due to gaps between plastic scintillator panels. We applied a data-driven method to estimate the muon mis-tag contribution in the signal region, as described in Ref.~\cite{COSINE-100:2018ged}. Because of the 4$\pi$ solid-angle coverage of the LS active shield~\cite{Adhikari:2020asl}, almost no events can reach the NaI(Tl) crystals without hits on the LS detector. Therefore, we do not consider the mis-tagged muon contribution for the single-hit events.

Thermal neutron capture by copper or iron nuclei in the shielding materials produces $\gamma$-rays with energies as high as 8\,MeV via ($n, \gamma$) reactions. 
The thermal neutron and total neutron flux  measured at the Yangyang underground laboratory are (1.44$\pm$0.15)$\times 10^{-5}$/(cm$^2 \cdot$s) and (4.46$\pm$0.66)$\times 10^{-5}$/(cm$^2 \cdot$s), respectively~\cite{Yoon:2021tkv}.
The neutron-induced events shown in Fig.~\ref{fig_event} (b) were simulated based on this flux.

In addition, we estimate the expected background events from $^{8}$B solar neutrino elastic scattering on electrons. 
Table~\ref{table-1} presents the expected backgrounds from the aforementioned contributions for the single-hit and multiple-hit channels, which are compared with the measured data. 
The measured data agree with the total expected backgrounds within their uncertainties. 

\begin{table*}
  \begin{center}
   \caption{The expected number of background events and the observed events from the 1.7 years COSINE-100 dataset are shown for both the single-hit and the multiple-hit channels.
   The individual contributions from environmental $\beta/\gamma$, thermal neutron capture, muon mis-tag, and solar neutrino are also listed.
	}
    \label{table-1}
      \begin{tabular}{c|c|c|c|c|c||c|c|c|c|c}
      \hline
			Energy		& \multicolumn{5}{c||}{Single-hit} & \multicolumn{5}{c}{Multiple-hit} \\
			\cline{2-11}
			(MeV) & $\beta/\gamma$ & neutron & neutrino & total & data & $\beta/\gamma$ & neutron & muon& total & data \\
      \hline
			4$-$6  & 172$\pm$26 & 203$\pm$30 & 0.039$\pm$0.006 & 375$\pm$40 & 322 & 12$\pm$2 & 889$\pm$91 & 15$\pm$2& 915$\pm$91 & 873 \\
			6$-$8  & 0 & 592$\pm$63 & 0.024$\pm$0.004 & 592$\pm$63 & 545 & 0 & 1165$\pm$120 & 16$\pm$2& 1181$\pm$120 & 1194 \\
			8$-$10 & 0 & 60$\pm$25 & 0.011$\pm$0.002 & 60$\pm$25 & 78 & 0 & 30$\pm$11 & 21$\pm$3& 51$\pm$12 & 37 \\
			$>$10  & 0 & 0          & 0.003$\pm$0.001 & 0.003$\pm$0.001          &  0  & 0 & 2$\pm$1      & 211$\pm$4 & 213$\pm$4      & 218       \\
      \hline
    \end{tabular}
  \end{center}
\end{table*}

In Ref.~\cite{Agashe:2014yua}, it is proposed that the boosted, lighter $\chi_1$ dark matter particles are produced in the pair-annihilation of two heavier $\chi_0$ with a total flux,
\begin{equation}
		\mathcal{F} = 1.6 \times 10^{-4} \mathrm{cm}^{-2} \mathrm{s}^{-1} \left(\frac{<\sigma v>_{0 \rightarrow 1}}{5 \times 10^{-26} \mathrm{cm}^{3} \mathrm{s}^{-1}} \right) \left(\frac{\mathrm{GeV}}{m_0}\right)^{2},
\label{flux}
\end{equation} 
where the reference value for $<\sigma v>_{0 \rightarrow 1}$, which is the velocity-averaged annihilation cross section of $\chi_0\chi_0 \rightarrow \chi_1 \chi_1$, corresponds to a correct dark matter thermal relic density for $\chi_0$ that is derived by a so-called ``assisted'' freeze-out mechanism~\cite{Belanger:2011ww}, and $m_0$ denotes the mass of $\chi_0$. This production rate is subject to uncertainties in the dark matter halo models~\cite{Navarro:1995iw,Kravtsov:1997dp,Moore:1999gc}. Here we assume the NFW halo profile~\cite{Navarro:1995iw,Navarro:1996gj} described in Ref.~\cite{Agashe:2014yua}. 
The relativistic $\chi_1$ (mass $m_1$) travels and interacts with terrestrial detector elements either elastically or inelastically.
We consider $\chi_1 e^-$ elastic scattering via a mediator $X$ (mass $m_X$) exchange.

We generate expected signals for various values of BDM parameters ($\gamma_1 = m_0/m_1$, $m_X$, and $\epsilon$, where $\epsilon$ is the coupling  between the dark sector mediator $X$ and the electron) based on Refs.~\cite{Kim:2016zjx,Giudice:2017zke}. The generated signal events undergo detector simulation~\cite{cosinebg,cosinebg2} and event selection. To search for BDM-induced events, we use a Bayesian approach with a likelihood function based on Poisson probability~\cite{COSINE-100:2021xqn}. We perform binned maximum likelihood fits to the measured energy spectra for two different channels of the single-hit and the multiple-hit events for each signal of the various BDM parameters. Each crystal for each channel is fitted with a crystal- and channel-specific background model and a crystal- and channel-correlated BDM signal for the combined fit by multiplying the ten likelihoods of the five crystals. We use evaluated background contributions to set Gaussian priors for the known background rates. 

\begin{figure*}[!htb] 
\begin{center}
		\begin{tabular}{cc}
\includegraphics[width=0.48\textwidth]{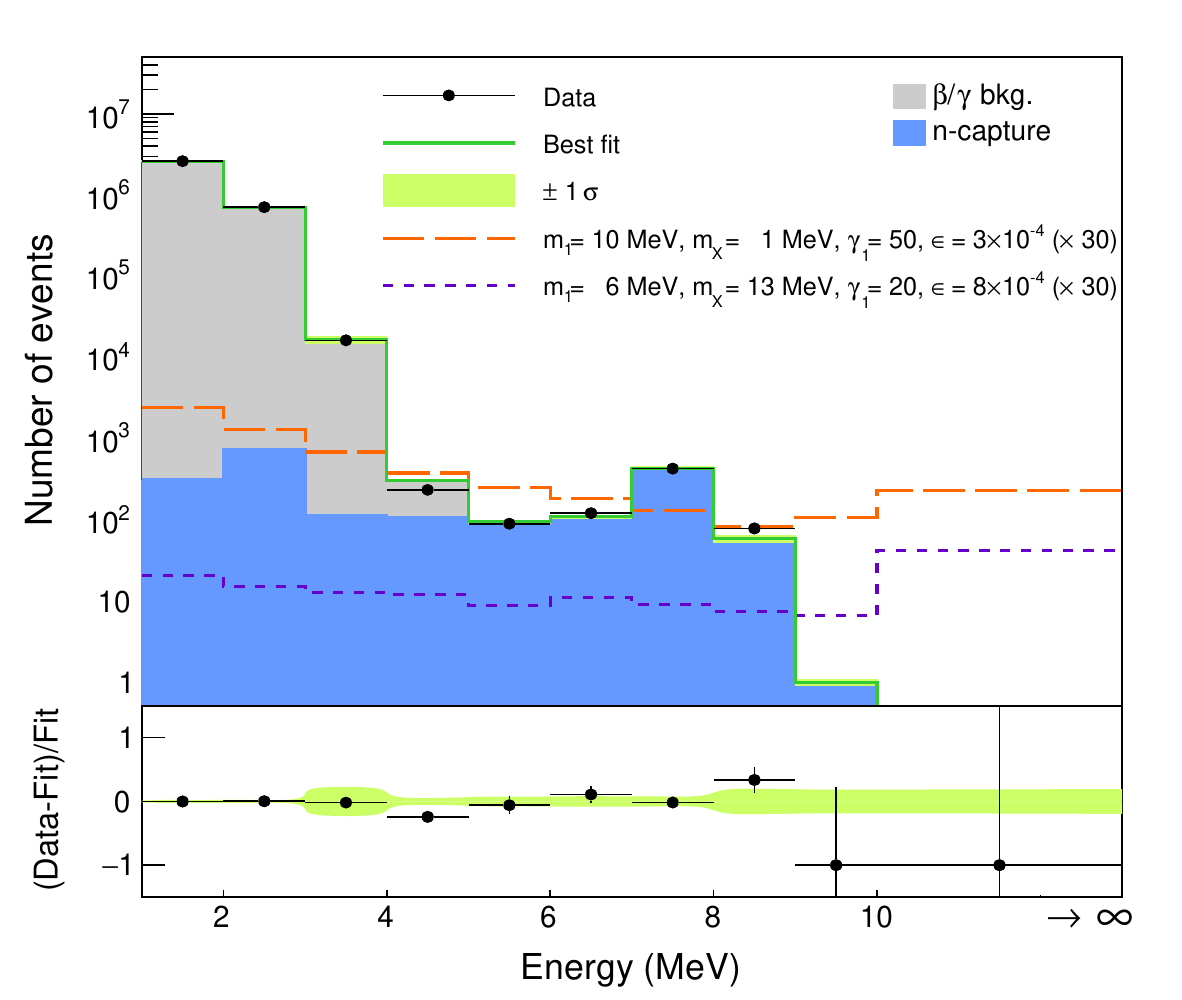}  &
\includegraphics[width=0.48\textwidth]{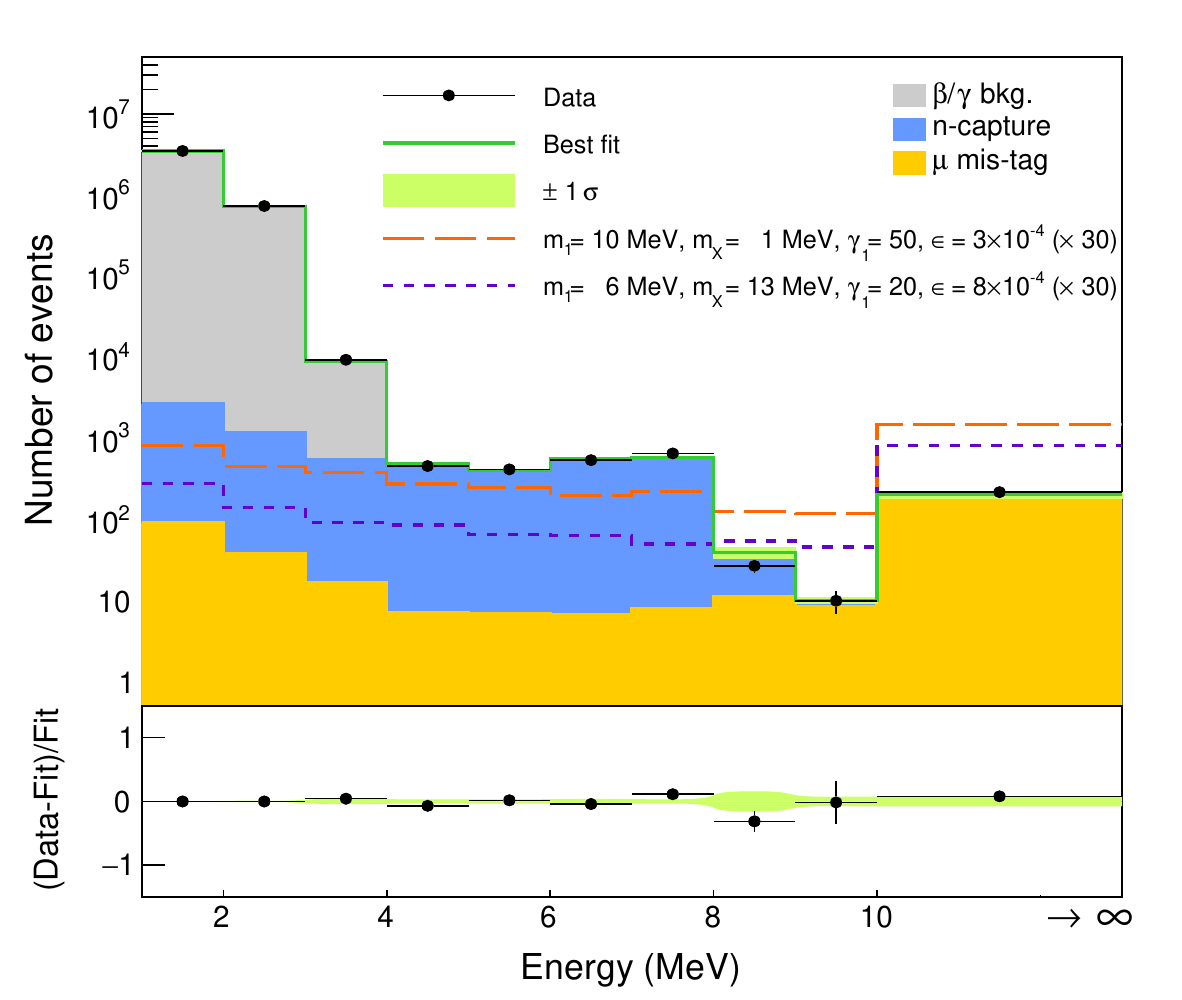}  \\
(a) single-hit data & (b) multiple-hit data \\
		\end{tabular}
\caption{
		The summed energy spectra for the five crystals (black filled circles) and the best fit (green solid lines) with BDM signal of $m_1$=6\,MeV, $m_X$=13\,MeV, $\gamma_1$=20, $\epsilon$=8$\times10^{-4}$ are presented for the single-hit events (a) and the multiple-hit events (b). Fitted contributions to the background from $\beta/\gamma$ radiation, neutron-capture, and muon mis-tag are indicated. The green bands are the 68\,\% confidence level (CL) intervals of the uncertainties obtained from the likelihood fit. 
		For presentation purposes, we draw the BDM signal shapes assuming BDM parameters of $m_1$=10\,MeV, $m_X$=1\,MeV, $\gamma_1$=50, $\epsilon$=3$\times10^{-4}$ and $m_1$=6\,MeV, $m_X$=13\,MeV, $\gamma_1$=20, $\epsilon$=8$\times10^{-4}$ with $\times$30 amplification of the signal amplitude. 
}
\label{fig_Espec}
\end{center}
\end{figure*}

Figure~\ref{fig_Espec} presents an example of the maximum likelihood fit for BDM signals with assumed parameters of  
$m_1$=6\,MeV, $m_X$=13\,MeV, $\gamma_1$=20, $\epsilon$=8$\times 10^{-4}$. 
The summed event spectra for the five crystals in the single-hit (a) and multiple-hit (b) events are shown together with the best-fit result. 
For comparison, the expected signals for the BDM parameters 
$m_1$=10\,MeV, $m_X$=1\,MeV, $\gamma_1$=50, $\epsilon$=3$\times 10^{-4}$ and $m_1$=6\,MeV, $m_X$=13\,MeV, $\gamma_1$=20, $\epsilon$=8$\times10^{-4}$ are presented. 
No excess of events that could be attributed to BDM interaction is found for the considered BDM signals. The posterior probabilities of signals are consistent with zero in all cases, and 90\,\% CL upper limits are determined. 

\begin{figure}[!htb] 
\begin{center}
\includegraphics[width=0.48\textwidth]{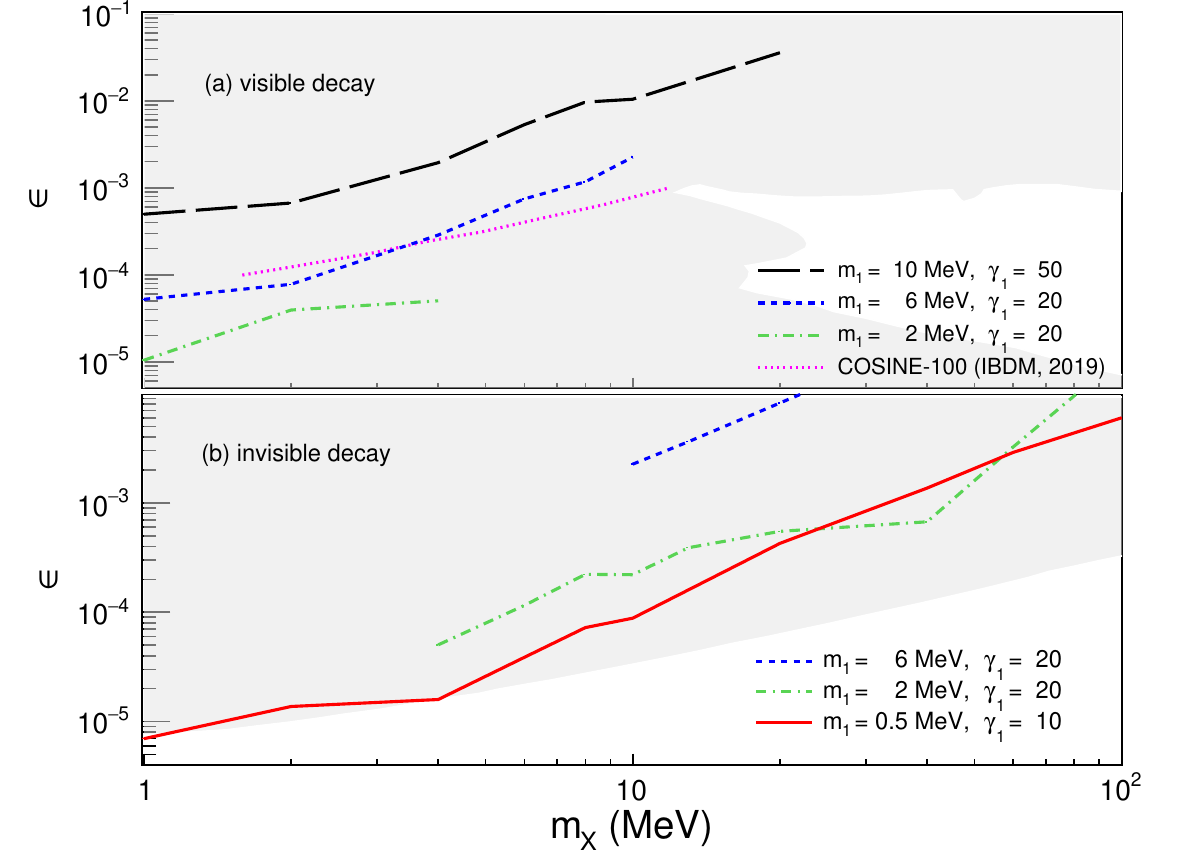}  
\caption{
		Interpretation of the BDM searches with 1.7 years of COSINE-100 data in terms of the dark photon mass (m$_X$) and coupling ($\epsilon$) parameters is presented. (a) When $m_X<2m_1$,  our limits  are compared with the currently excluded parameter space (shaded region) for the visible decay mode, including results from E141~\cite{Riordan:1987aw}, NA48~\cite{Batley:2015lha}, NA64~\cite{NA64:2019auh}, Babar~\cite{Lees:2014xha}, as well as bounds from the electron anomalous magnetic moment~\cite{Davoudiasl:2014kua} and our previous IBDM search with 59.5\,days of COSINE-100 data~\cite{COSINE-100:2018ged}. (b) When $m_X\ge2m_1$, our limits are compared with the currently excluded parameter space (shaded region) for the invisible decay mode from NA64~\cite{Andreev:2021fzd} and Babar~\cite{BaBar:2017tiz} experiments. 
}
\label{fig_result}
\end{center}
\end{figure}

We interpret this result in the context of dark photon phenomenology by assuming that the interaction between the standard model particles and the dark sector particles is mediated by a dark photon. It allows us to compare this result with other dark photon searches in terms of the parameters $m_X$ and $\epsilon$.  A similar interpretation with 59.6\,days of COSINE-100 data for IBDM was presented in Ref.~\cite{COSINE-100:2018ged}. In our analysis, we generate signals using different sets of model parameters, fixing $m_1$ and $\gamma_1$ while varying $m_X$. Figure~\ref{fig_result} shows the measured 90\,\% CL upper limits obtained from the 1.7 years of COSINE-100 data for the aforementioned model parameters. 
We compare our results with those of direct dark photon searches for both the visible decay mode ($m_X<2m_1$) and the invisible decay mode ($m_X\ge2m_1$) in Fig.~\ref{fig_result} (a) and (b)\footnote{Note that additional constraints from cosmological and astrophysical observations, depending on the detailed model of the dark sector particles discussed in Ref.~\cite{Kamada:2021muh}, need to be taken into account.}, respectively. Notably, for the invisible mode, our analysis yields a  competitive limit for the dark photon mass below 4\,MeV, assuming parameters of  $m_1$=0.5\,MeV and $\gamma_1$=10. This result highlights the complementarity of our search for the dark photon, although the specific model discussed in this paper has to be assumed.

\begin{figure}[!htb] 
\begin{center}
\includegraphics[width=0.48\textwidth]{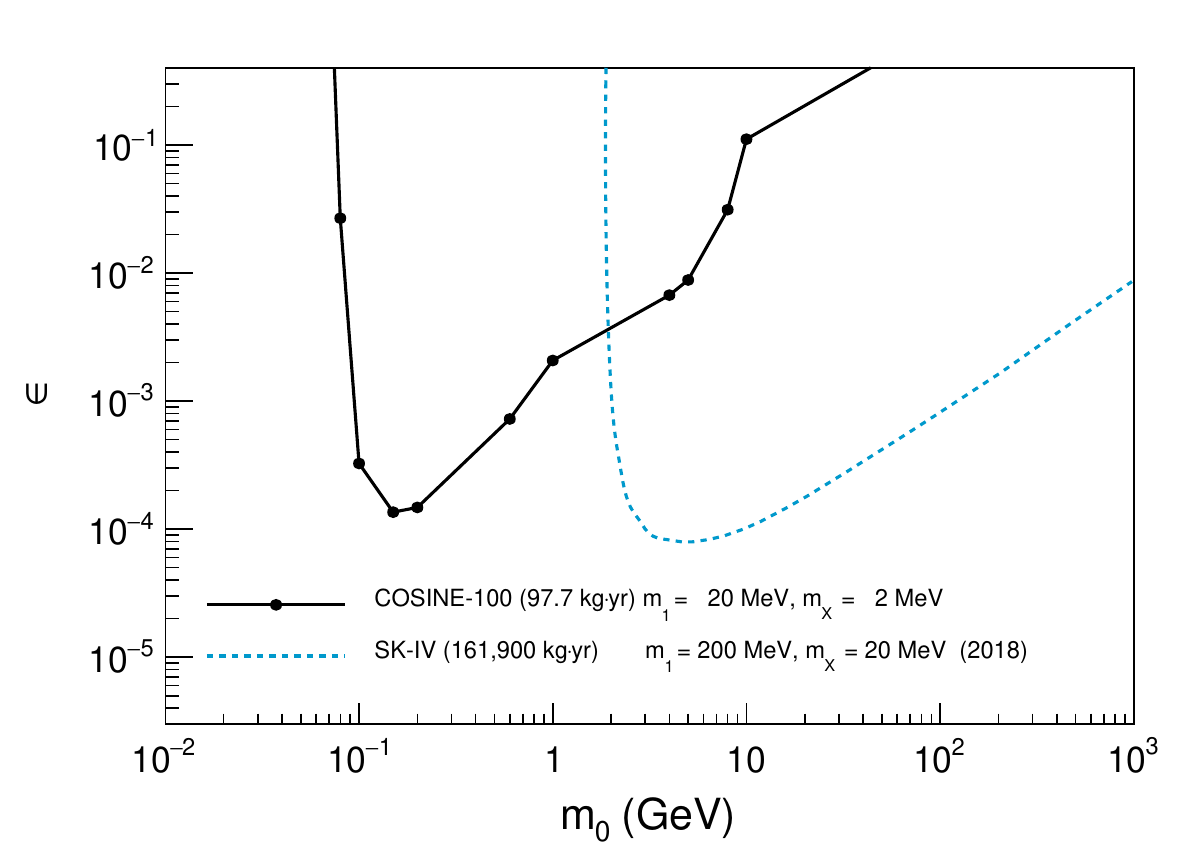} 
\caption{
		 Interpretation of the BDM searches with 97.7\,kg$\cdot$years of COSINE-100 data in terms of the relic dark matter mass (m$_0$) and $\epsilon$ parameters is conducted, assuming $m_1$=20\,MeV and $m_X$=2\,MeV. This result is then compared with limits from 161.9\,kiloton$\cdot$years of SK IV data, assuming $m_1$=200\,MeV and $m_X$=20\,MeV~\cite{Super-Kamiokande:2017dch}. To ensure a proper comparison with the SK result, we assume the coupling constant between $\chi_1$ and $X$ through elastic scattering to be 0.5. 
}
\label{fig_result2}
\end{center}
\end{figure}

An additional interpretation of the Super-Kamiokande~(SK) IV search result~\cite{Super-Kamiokande:2017dch} considers the relic dark matter mass ($m_0$) and coupling ($\epsilon$) parameter space, as shown in Fig.~\ref{fig_result2}. Two results are obtained using the same NFW halo profile~\cite{Navarro:1995iw} with a 0.5 coupling constant between the BDM and the mediator through the elastic interaction.  
Despite using a much smaller dataset of 97.7\,kg$\cdot$years compared to the 161.9\,kiloton$\cdot$years SK exposure, the lowest bound for $\epsilon$ at $m_0$ of about 200\,MeV is at a similar level as that of the SK search result for an $m_0$ of about 5\,GeV. 
Because of the well understood  backgrounds in the COSINE-100 detector above 4\,MeV, the COSINE-100 data is complementary to results from the SK detector in searching unexplored parameter space at low-mass dark matter. 

In summary, we searched for evidence of boosted dark matter (BDM) by observing energetic electron recoil events induced by the elastic scattering of the BDM. Based on  1.7 years of COSINE-100 data, we found no evidence of BDM interaction, and we set 90\,\% CL limits for various model parameters. Our investigation of dark photon interactions explored a parameter space that complements  other dark photon search experiments. 
We also demonstrate that a small-scale dark matter search detector has some unique advantages for the low-mass dark matter in the BDM scenario compared to the much larger neutrino detectors. 
Although our results are interpreted in the context of the BDM model that elastically scatters electron, this search can apply to any theory that predicts an excess of events in electron recoil  of a few MeV, for which the COSINE-100 detector has world-competitive sensitivity.

\acknowledgments
We thank Jong-Chul Park and Seodong Shin for insightful discussions. We thank the Korea Hydro and Nuclear Power (KHNP) Company for providing underground laboratory space at Yangyang and the IBS Research Solution Center (RSC) for providing high performance computing resources.  This work is supported by:  the Institute for Basic Science (IBS) under project code IBS-R016-A1, NRF-2021R1A2C3010989 and NRF-2021R1A2C1013761, Republic of Korea; NSF Grants No. PHY-1913742, DGE-1122492, WIPAC, the Wisconsin Alumni Research Foundation, United States; STFC Grant ST/N000277/1 and ST/K001337/1, United Kingdom; Grant No. 2021/06743-1 and 2022/12002-7 FAPESP, CAPES Finance Code 001, CNPq 131152/2020-3 and 303122/2020-0, Brazil.
\bibliographystyle{PRTitle}
\providecommand{\href}[2]{#2}\begingroup\raggedright\endgroup
\end{document}